\documentclass[]{aa}

\usepackage{amsmath}
\usepackage{amssymb}
\usepackage{url}
\usepackage{graphicx}
\usepackage{times}
\usepackage[utf8]{inputenc}
\usepackage{xspace}
\usepackage{color}
\usepackage{multirow}
\usepackage{multicol}
\usepackage{float}

\bibpunct{(}{)}{;}{a}{}{,}

\usepackage{etoolbox}
\makeatletter
\patchcmd\@combinedblfloats{\box\@outputbox}{\unvbox\@outputbox}{}{%
  \errmessage{\noexpand\@combinedblfloats could not be patched}%
}%
\makeatother

\def\apjl{ApJL}
\def\apjs{ApJSS}

\def\aap{A\&A}

\def\tv{\theta_{\rm v}}
\def\tj{\theta_{\rm jet}}
\def\ie{i.e.\xspace}
\def\eg{e.g.\xspace}

\newcommand{\LEt}[1]{\iffalse {#1} \fi} 

\def\grbflux{$3.7\pm 0.9\rm\,ph\,cm^{-2}s^{-1}$\xspace}
\def\grbfluence{$\left(2.8\pm 0.2\right)\times 10^{-7}\rm\,erg\,cm^{-2}$\xspace}
\def\grbNTfluence{$\left(1.8\pm 0.4\right)\times 10^{-7}\rm\,erg\,cm^{-2}$\xspace}
\def\grbBBfluence{$\left(0.61\pm 0.12\right)\times 10^{-7}\rm\,erg\,cm^{-2}$\xspace}
\def\grbtime{2017-08-17 12:41:06\xspace}
\def\grbduration{$T_{90}=2.0\pm 0.5\,\rm s$\xspace}
\def\grbdelay{$1.74\,\rm s$\xspace}
\def\grbepeak{$E_{\rm peak} = 185 \pm 62\,\rm keV$\xspace}

\def\grbtemp{${\rm k_B}T=10.3\pm 1.5\,\rm keV$\xspace}

\def\E0fit{$E_0=(3.1^{+2.1}_{-1.3})\times 10^{49}\,{\rm erg}$}

\def\R0fit{$R_0=54^{+50}_{-25}\,{\rm km}$}
\def\T0fit{$T_0=(2.6^{+1.5}_{-1.0})\times 10^{11}\,{\rm K}$}

\begin{document}

\title{Interpreting GRB170817A as a giant flare from a jet-less double neutron star merger}
\titlerunning{GRB170817A as a giant flare from a NS-NS merger}
\author{O.~S.~Salafia\inst{\ref{unimib},\ref{oab.me},\ref{infn.mib}}\thanks{E--mail: omsharan.salafia@brera.inaf.it} \and G.~Ghisellini\inst{\ref{oab.me}} \and G.~Ghirlanda\inst{\ref{oab.me}} \and M.~Colpi\inst{\ref{unimib},\ref{infn.mib}}}

\institute{Università degli Studi di Milano-Bicocca, Dip. di Fisica ``G. Occhialini'', Piazza della Scienza 3, I-20126 Milano, Italy\label{unimib} \and INAF -- Osservatorio Astronomico di Brera, via E. Bianchi 46, I-23807 Merate, Italy\label{oab.me} \and INFN -- Sezione di Milano-Bicocca, Piazza della Scienza 3, I-20126 Milano, Italy\label{infn.mib}}

\authorrunning{O.~S.~Salafia et al.}

\date{Received: November 2017 / Accepted: August 2018}

\abstract{
We show that the delay between GRB170817A and GW170817 is incompatible with de-beamed emission from an off-axis relativistic jet. The prompt emission and the subsequent radio and X-ray observations can instead be interpreted within a giant-flare-like scenario, being the result of a relativistic outflow driven by the ultra-strong magnetic field produced by magnetohydrodynamic amplification during the merger of the progenitor double neutron-star binary. Within such a picture, the data indicate that the outflow must be endowed with a steep velocity profile, with  a relatively fast tail extending to $\Gamma\sim 8$. Since the conditions for the launch of such an outflow are relatively general, and the presence of a velocity profile is a natural expectation of the acceleration process, most neutron star binary mergers should feature this quasi-isotropic, hard X-ray emission component, that could be a powerful guide to the discovery of additional kilonovae associated to relatively nearby gravitational wave events.
}


\maketitle

\section{Introduction}

Before 17 August 2017, events like GRB170817A have probably been detected several times by instruments such as \textit{Fermi}/GBM and possibly \textit{Swift}/BAT without attracting too much attention. Due to the low flux and fluence, this kind of burst usually does not become a candidate for a follow-up aimed at searching for the afterglow and identifying the host galaxy. These events simply end up populating the highly incomplete part of the ``$\log N-\log S$'' (the fluence distribution of the sample), and are usually taken out of the flux-limited samples used for population studies.

The association with the gravitational wave (GW) event GW170817 reveals, however, that at least some of these low-flux events might offer extremely precious information about one of the most interesting astrophysical events: a binary neutron star (NS-NS) merger \citep{Abbott2017,Abbott2017b,Abbott2017a}.

The fact that such an association entered the scene so early is astonishing from many points of view: even though NS-NS mergers have been among the best candidate short gamma-ray burst (SGRB) progenitors for a long time now  \citep{Eichler1989}, the most recent predictions based on available observational data \citep{Ghirlanda2016,Wanderman2014a} indicate a very small probability for the detection of an SGRB located within the Advanced LIGO/Virgo network range during the first and second run. Moreover, all measurements of SGRB half-opening angles to date \citep{Soderberg2006a,NicuesaGuelbenzu2011,Fong2012,Fong2013,Troja2016} point to narrow jets ($\tj\lesssim 10^\circ$ -- even though serious selection effects could be at play), implying that the probability of an on-axis or slightly off-axis jet associated with the very first GW from a NS-NS merger is very small. Last but not least, several studies \citep[\eg][]{Ruiz2017,Murguia-Berthier2016,Margalit2015} seem to indicate that not all NS-NS mergers are capable of producing a jet, making the association even more unlikely.

It could certainly be the case that all or some of the above expectations and prejudices about SGRB jets are simply incorrect. On the other hand, many features of GRB170817A suggest quite naturally that it does not belong to the SGRB population we are used to. Its isotropic equivalent energy is several orders of magnitude below the least energetic SGRB known so far, despite the spectral peak energy being only moderately low with respect to the known population \citep{Nava2011,Zhang2017}. Indeed, these facts have been taken by many as hints that GRB170817A is an ordinary or structured SGRB jet seen off-axis \citep[\eg][]{Pian2017,Ioka2017,Burgess2017,Zhang2017,Lamb2017,He2017,Kathirgamaraju2017}, while others interpret GRB170817A as emission from the jet cocoon \citep[\eg][]{Kasliwal2017,Bromberg2017,Piro2017,Gottlieb2017a,Lazzati2017}. 

In this work, we explore whether GRB170817A can be interpreted as emission from an isotropic fireball powered by the strong magnetic field produced during the progenitor NS-NS merger, as already proposed in \citealt{Salafia2017b} (Paper I hereafter). 

\section{Observational properties of GRB170817A}\label{sec:obs_properties}
The GRB triggered \textit{Fermi}/GBM on  \LEt{please see AandA language guide Sect. 1.2 for accepted date formats and edit throughout.}\grbtime, just \grbdelay after the estimated merger time of GW170817 \citep{Abbott2017a}. A corresponding signal was also detected by \textit{INTEGRAL}/SPI-ACS \citep{Savchenko2017}, providing important confirmation. The duration of the burst has been estimated as \grbduration. The 64 ms peak flux in the 10--1000 keV band was \grbflux, and the fluence was \grbfluence. Detailed analysis \citep{Goldstein2017,Zhang2017} indicates the possible presence of two components: a non-thermal component  dominating the early part of the light curve, whose spectrum is fit by a power-law with an exponential cut-off with \grbepeak, and has a fluence of \grbNTfluence; and a thermal component visible in the tail of the light curve, which can be fit by a blackbody with \grbtemp (where $\rm k_B$ is the Boltzmann's constant) and a fluence  of \grbBBfluence, which corresponds to an isotropic equivalent energy $(1.20\pm 0.23)\times 10^{46}\,{\rm erg}$ at $d_{\rm L}\approx 40\,{\rm Mpc}$ (\ie the distance to the host galaxy NGC4993, \citealt{Hjorth2017,Im2017}). According to \citet{Goldstein2017}, the thermal component could be present since the beginning, masked initially by the non-thermal emission.
Multi-wavelength follow-up of the event \citep{Abbott2017a} resulted in several important detections. In the ultraviolet (UV), optical, and near infrared (NIR) during the week following the merger, a relatively bright optical transient was discovered \citep{Coulter2017,Valenti2017} and extensively observed \citep[\eg][]{Andreoni2017,Arcavi2017,Chornock2017,Covino2017,Diaz2017,Drout2017,Evans2017,Hallinan2017,Pian2017,Pozanenko2017,Smartt2017,Utsumi2017}. Its nature is established \citep[\eg][]{Pian2017,Cowperthwaite2017,Gall2017,Kilpatrick2017,McCully2017,Nicholl2017,Smartt2017} as being nuclear-decay-powered emission from the expanding NS-NS merger ejecta (\ie a \textit{kilonova}, \citealt{Li1998,Metzger2010,Metzger2016}).

Another transient, which we interpret as a GRB-related afterglow, has been detected in X-rays by \textit{Chandra} \citep{Troja2017,Margutti2017,Haggard2017,Troja2018,Margutti2018} and \textit{XMM-Newton} \citep{DAvanzo2018} in the Optical by \textit{HST} \citep{Lyman2018,Margutti2018} and in radio by several facilities \citep{Hallinan2017,Mooley2017,Margutti2018,Dobie2018}. 
Observations between 9 and 150 days after GRB170817A show a spectrum which is consistent with a single power law extending from a few gigahertz to the X-ray band \citep{Margutti2018}. The luminosity during that period kept rising approximately as $t^{0.8}$ \citep{Mooley2017}. Observations in the  optical \citep{Margutti2018}, X-rays \citep{DAvanzo2018} and radio \citep{Dobie2018} around 150 days started to show some signs of a flattening in the light curve, and a peak at approximately 160 days was later confirmed \citep{Alexander2018}. When interpreted as synchrotron emission from a blast wave in a constant-density interstellar medium (ISM), these observations indicate \citep{Nakar2018} that the emitting material is mildly relativistic ($\Gamma\sim\mathrm{a\,few}$).

\section{The time delay between GW170817 and GRB170817A: relatively short for an off-axis jet}\label{sec:time_delay}
 
As discussed in \citet{Salafia2016} and \citet{Murguia-Berthier2017} for example, the observed duration of a GRB should not depend on the viewing angle. Individual pulses that constitute the light curve become longer with increasing viewing angle, but their separation remains unchanged. The result is a smoother pulse shape when the jet is seen off-axis, but without significant change in the total duration (provided that single pulses are much shorter than the total duration of the burst).

The delay with respect to the jet launch time (and thus the merger time in our case), instead, increases with the viewing angle. We show in what follows that it is hard to reconcile a delay of $\sim$2 seconds with emission from a jet seen under a large viewing angle \citep[somewhat similar arguments are outlined in][]{Shoemaker2017}. 

\begin{figure}
 \begin{center}
 \includegraphics[width=0.75\columnwidth]{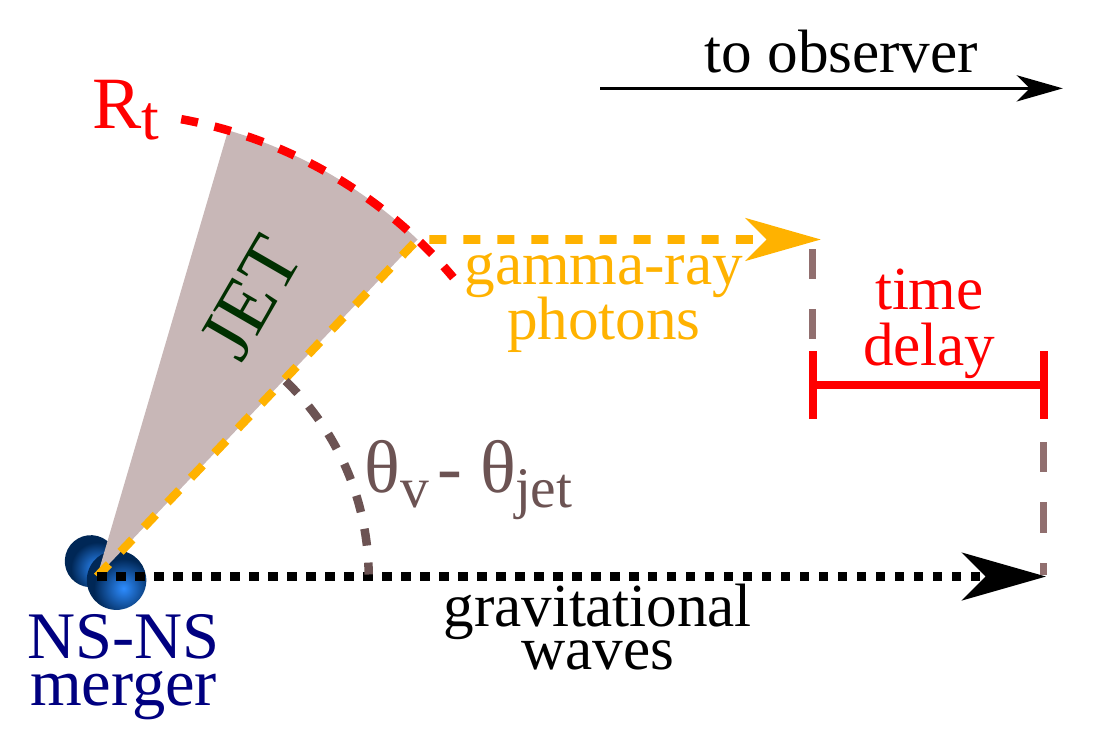} 
 \end{center}
 \caption{\label{fig:t_delay} If the gamma-ray prompt emission is due to an off-axis jet, the arrival time difference between the last gravitational waves and the first gamma-ray photons is dominated by the time it takes for the jet to become transparent in the observer frame.}
\end{figure}

In order to emit gamma-rays, the jet must expand enough to become transparent.
We estimate
the transparency radius $R_t$ as \citep[\eg][]{Daigne2002}
\begin{equation}
 R_t = \frac{L_{\rm K,iso} \sigma_T}{8\pi m_p c^3 \Gamma^3}
 \label{eq:Rph_jet}
,\end{equation}
where $\sigma_T$ is the Thomson cross-section, $m_p$ is the proton mass, and we are assuming an electron fraction of unity. $L_{\rm K,iso}$ in this expression is the isotropic equivalent kinetic luminosity of the outflow, that is, $L_{\rm K,iso}=\Gamma\dot{M}_{\rm iso}c^2$. If we assume that the jet is launched a short time ($\ll 1\,{\rm s}$) after the merger, the arrival time difference between the latest gravitational waves and the first photons is
\begin{equation}
 t_\gamma - t_{\rm GW} \approx \frac{R_t}{\beta c}\left(1-\beta \cos\left(\tv-\tj\right)\right) 
.\end{equation}
This accounts for the fact that fluid elements on the jet border must travel up to $R_t$ at a speed of $\beta c$ before being able to emit the gamma-ray photons (see Fig.~\ref{fig:t_delay}).

\begin{figure}
 \begin{center}
 \includegraphics[width=\columnwidth]{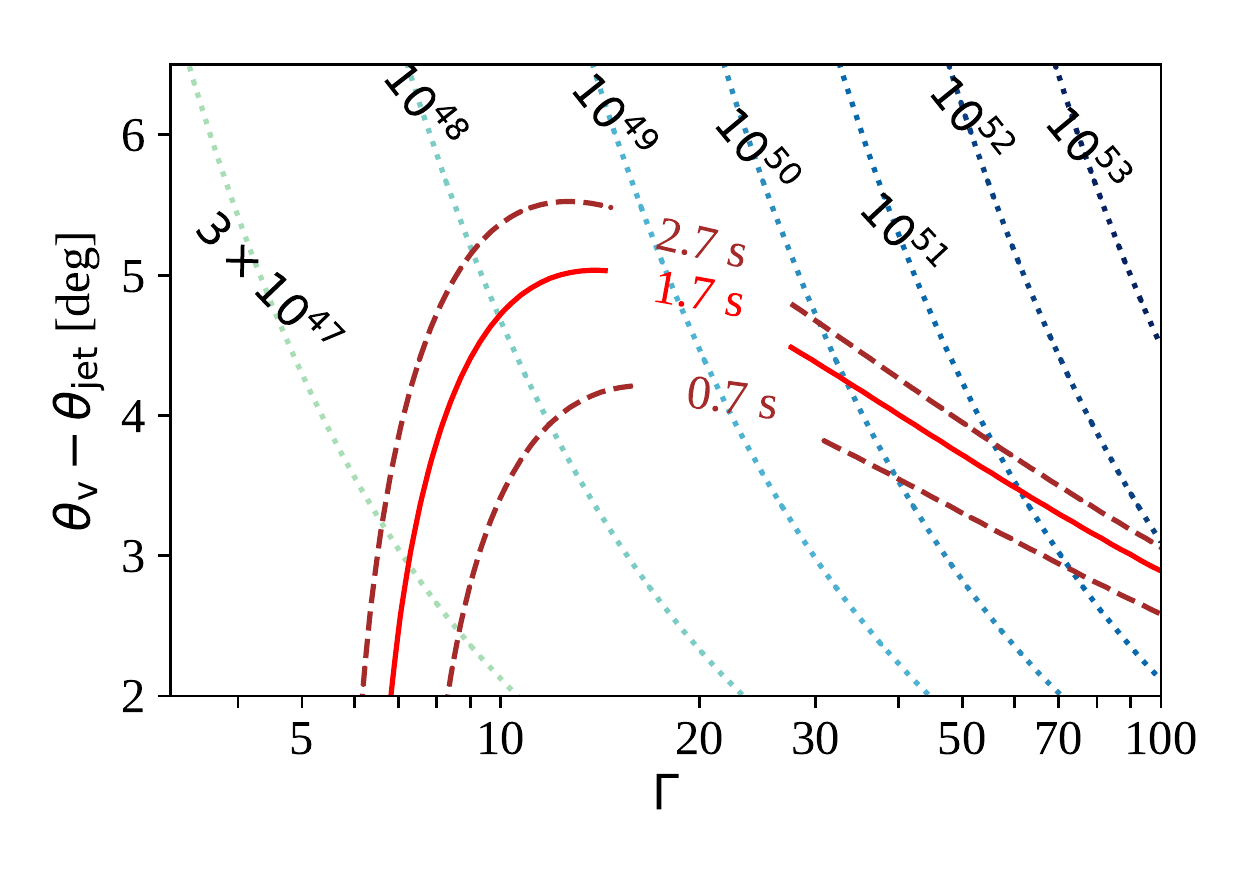} 
 \end{center}
 \caption{\label{fig:delay_angle_gamma}  The solid red and dashed brown lines represent contours of the gamma-ray photon arrival time delay with respect to the last gravitational waves for a given jet bulk Lorentz factor $\Gamma$ and off-axis viewing angle $\tv-\tj$. The dotted lines represent contours of the on-axis jet luminosity $L_{\rm iso}(0)$ corresponding to an off-axis luminosity of $L_{\rm iso}(\tv)=10^{47}\,{\rm erg/s}$.}
\end{figure}

We can relate the kinetic luminosity of the jet to the on-axis gamma-ray luminosity by assuming that 10\%\ of the kinetic luminosity is converted into photons, that is, $L_{\rm iso}(\tv=0) = 0.1\,L_{\rm K,iso}$ \citep[as in \eg][see also \citealt{Beniamini2015} who show that this is a typical conversion efficiency]{Kathirgamaraju2017}. We compute the corresponding off-axis luminosity $L_{\rm iso}(\tv)$ using Eq.~11 of \citet{Salafia2016}. If we require that $L_{\rm iso}(\tv)= 10^{47}\,{\rm erg/s}$ as in GRB170817A \citep{Goldstein2017}, and we assume a jet half-opening angle $\tj$, we can compute the corresponding on-axis luminosity $L_{\rm iso}(0)$ and arrival time delay $t_{\gamma}-t_{\rm GW}$ for various combinations of the bulk Lorentz factor $\Gamma$ and off-axis viewing angle $\tv-\tj$. Figure~\ref{fig:delay_angle_gamma} shows the contours of these quantities for $\tj=0.2\,{\rm rad}=11.5^\circ$ (the figure would be very similar for $\tj=0.1\,{\rm rad}$ or $\tj=0.3\,{\rm rad}$). The red solid contour corresponds to the actual $1.7\,{\rm s}$ delay time as observed in GRB170817A, while the dotted contours represent the on-axis luminosities. 
The figure shows that a ``standard'' jet with a large Lorentz factor $\Gamma\gtrsim 70$ and an on-axis luminosity $L_{\rm iso}(0)\sim 10^{51}\,{\rm erg/s}$ is formally compatible with the observed time delay, but it requires fine-tuning of the viewing angle, which implies an extremely small probability. Moreover, the line of sight
being only slightly off the jet border, the jet afterglow should have turned very bright after a few days, which would conflict with the stringent early X-ray upper limits set by \textit{Swift} and \textit{NuSTAR} during the first week after GW170817 \citep[see \eg][]{Evans2017}. A `slow' fireball with a Lorentz factor around $\Gamma\lesssim 7$ and a much smaller energy, instead, would give the correct time delay both if seen on-axis or a few degrees off-axis. For these reasons, we believe that GRB170817A is more naturally explained by slow material moving towards the observer (which is the case both for the jet cocoon -- \citealt{Lazzati2016,Lazzati2017} -- and for the isotropic giant flare described in Paper I) than by a fast off-axis jet. The cocoon explanation, on the other hand, still requires the presence of a jet. Since the launch of a jet in the post-merger phase is not guaranteed \citep[\eg][]{Ruiz2017,Murguia-Berthier2016}, while the conditions leading to the giant flare seem to be typical for double neutron-star mergers \citep{Zrake2013,Giacomazzo2015}, the latter seems to be the most natural, and below we show that indeed both the prompt and the afterglow emission of GRB170817A can be understood within such a jet-less picture. Future observations of double neutron-star mergers and their electromagnetic counterpart will certainly provide the necessary evidence to discern between the possible scenarios outlined here and help to assess their relative frequency.

\section{Interpreting the GRB as an isotropic fireball}

\subsection{The scenario}\label{sec:scenario}
Let us briefly summarise the scenario described in Paper I: 
%
during the inital phase of the merger, the magnetic field is quickly amplified to $B\sim 10^{16}\,{\rm G}$ \citep{Price2006,Zrake2013,Giacomazzo2015}. If such a strong magnetic field transfers part of its energy to a small amount of plasma surrounding the merger, for example by reconnection or by impulsive magnetic acceleration \citep{Contopoulos1995, Granot2010, Lyutikov2011}, a relativistic fireball can be produced. 
The fireball, whose initial radius is $R_0$, promptly expands and accelerates to relativistic velocities under its own pressure\footnote{We refer here to the standard internal-energy-driven fireball acceleration scenario. The conclusions remain essentially unchanged if impulsive magnetic acceleration is considered as the acceleration mechanism.}. The acceleration ends when most of the initial internal energy $E_0$ has been transformed into kinetic energy, at a radius (we employ the usual notation $Q_x \equiv Q/10^x$ in \LEt{Please spell out all acronyms the first time they
appear in the paper, followed by the abbreviation in parentheses, both in
the abstract and again in the main text. After that, please only use the
abbreviation. See A and A language guide Section 5.2.4 www.aanda.org/language-editing}cgs units),
\begin{equation}
 R_{\rm a} \sim \Gamma R_0 = 10 \Gamma_1 R_0
.\end{equation}
The thickness of the fireball at this point is of the order of the initial radius $R_0$, but its intrinsic velocity profile \citep[developed during the initial acceleration phase, \eg][]{Piran1993} causes it to spread\footnote{This spreading was overlooked in Paper I.} as the expansion proceeds. The effect becomes significant only beyond a ``spreading'' radius $R_\mathrm{s}\sim \Gamma^2 R_0$ \citep[\eg][]{Meszaros1993}, after which the fireball thickness is given by $R/\Gamma^2$.
The opacity $\kappa$ of the fireball to high-energy photons is dominated by Thomson scattering, and thus 
\begin{equation}
 \kappa = Y_e \frac{\sigma_T}{m_p} = 0.4\,Y_e\,{\rm cm^2\,g^{-1}}
,\end{equation}
where $Y_e = n_e/(n_p+n_n)$ is the electron fraction. In what follows, we take $Y_e=0.5$, which is expected for relativistic, pair-dominated outflows \citep{Beloborodov2003}.
The photospheric radius of a thin shell is given by \citep[e.g.][]{Meszaros2006}
\begin{equation}
 R_{\rm ph} = \left(\frac{M \kappa}{4\pi}\right)^{1/2} \approx 4.2\times 10^{12}\,\Gamma_{1}^{-3/2} E_{0,49}^{1/2}\,\mathrm{cm}
 \label{eq:Rph}
.\end{equation}
Once the fireball becomes transparent, relic photons from the initial radiation-dominated phase are liberated. Assuming the energy to be initially completely thermal and confined in a spherical volume of radius $R_0$, the  photospheric temperature can be estimated as
\begin{equation}
 k_{\rm B}T_{\rm ph} =  (1+\beta)\Gamma^{4/3} k_{\rm B}T_0 \frac{R_0}{R_{\rm ph}} \approx 0.1\,\Gamma_{1}^{11/6}E_{0,49}^{-1/4}R_{0,6}^{1/4}\,{\rm keV}
 \label{eq:TBB}
,\end{equation}
where $\beta = (1-\Gamma^{-2})^{1/2}\approx 1$, producing a thermal pulse containing an energy of
\begin{equation}
 E_{\rm ph} =  \Gamma^{4/3} E_0 \frac{R_0}{R_{\mathrm{t}}} \approx 5.1\times 10^{43} \Gamma_{1}^{11/6}E_{0,49}^{1/2}R_{0,6}\,{\rm erg}
 \label{eq:EBB}
.\end{equation}
The duration of the pulse is approximately given by the angular timescale \citep[\eg][]{Salafia2016}
\begin{equation}
 t_{\rm ang} = \frac{R_{\rm ph}}{\Gamma^2 c} \approx 1.4\,\Gamma_{1}^{-5/2}E_{0,49}^{1/2}\,{\rm s}
 \label{eq:tang}
.\end{equation}
The above timescale is equal to the transverse timescale \citep[\eg][]{piran-physics_of_grbs04}, corresponding to the delay in the observer frame between the fireball formation and the arrival time of the first photons, which in our case should be equal to the time delay between GW170817 and GRB170817A.

As discussed in Paper I, dissipation of kinetic or magnetic energy within the outflow can give rise to additional emission, just as in gamma-ray bursts. 

\subsubsection{Constraints from observations}
A first constraint to the simple model outlined above is set by the $\grbdelay$ delay between the GW chirp and the start of the GRB.  Taking $\Gamma_1\lesssim 1$ as suggested by the arguments in Sect. \ref{sec:time_delay}, and setting $t_\mathrm{ang}=1.74\,\mathrm{s}$ in Eq.~\ref{eq:tang}, we have
\begin{equation}
 E_\mathrm{0,49}\lesssim 1.5
 \label{eq:energy_upper_limit}
.\end{equation}
The initial energy release in our scenario must happen in the vicinity of the merging neutron stars, for example within the light cone of the system, which is located at $\sim 100\,\mathrm{km}$ for a system rotating at an angular frequency of $\Omega=3000\,\mathrm{rad/s}$. We then see from Eqs.~\ref{eq:EBB} and \ref{eq:TBB} that the energy content of GRB 170817A and its temperature are not fully explained by photospheric emission, and we therefore need some form of energy dissipation in the outflow to explain the GRB in our scenario. On the other hand, if afterglow observations are to be interpreted as synchrotron emission from the external shock that forms as the shell decelerates in the ISM, the energy in the ejecta (assumed isotropic) is of the order of
$\sim 10^{50}\,\mathrm{erg}$ \citep{Nakar2018,Huang2018}, meaning that inequality (\ref{eq:energy_upper_limit}) would be violated, that is,~the shell was still optically thick at the time at which it should have produced the GRB. As we show in the following section, this seemingly unsolvable tension can be addressed differently if the ejecta velocity profile is taken into account.

\subsection{The velocity profile}\label{sec:profiles}
Several authors \citep[\eg][]{Vitello1976,Piran1993,Meszaros1993,Bisnovatyi-Kogan1995a} studied the hydrodynamics of an expanding relativistic fireball in spherical symmetry. Asymptotic scalings of the main quantities are established, but the details of the profile are only accessible by numerical integration of the relativistic Euler equations. Nevertheless it seems clear that, regardless of the initial conditions, during its initial accelerated expansion the fireball undergoes a rearrangement and develops a velocity profile \citep{Piran1993}, with a Lorentz factor spread $\Delta\Gamma\sim \Gamma$ across the pulse \citep{Piran1993,Meszaros1993}. The same holds also if the shell is launched by impulsive magnetic acceleration \citep{Contopoulos1995,Granot2010,Lyutikov2011}. The presence of such a profile is indeed a natural outcome of the shell acceleration. The transparency condition in this case becomes simply
\begin{equation}
 R_\mathrm{ph}(\Gamma)\approx \left(\frac{M(>\Gamma)\kappa}{4\pi}\right)^{1/2}
,\end{equation}
where $M(>\Gamma)$ is the mass in ejecta moving with a Lorentz factor larger than $\Gamma$. This is usually neglected in GRBs because the time (in the observer frame) it takes for the photosphere to recede from $2\Gamma$ to $\Gamma$ is small if $\Gamma\gtrsim 100$, but in our mildly relativistic case this effect can make the difference. The idea is therefore that the prompt emission is produced at a time when only a fraction of the ejecta is transparent. Let us assume that the kinetic energy dissipation in the outflow, independently from its nature, has an efficiency of $\eta\lesssim 0.1$, that is,~only a relatively small fraction of the ejecta kinetic energy is converted into prompt radiation. We therefore require that the fraction of the ejecta that is transparent at an observer time \grbdelay contains an energy larger than $\eta^{-1}$ times the GRB energy, namely $E(>\Gamma)\gtrsim E_\mathrm{iso}/\eta = 6\times 10^{47}\,\mathrm{erg}$. Several authors \citep[e.g.][]{Mooley2017,Huang2018} have shown that a power-law energy profile $E(>\Gamma)=E_\star (\Gamma\beta)^{-\alpha}$ with $\alpha\sim 5$ -- $7$ is consistent with the observed afterglow, so let us adopt this parametrization. We have $M(>\Gamma)\sim E(>\Gamma)/\Gamma c^2$, and therefore
\begin{equation}
 t_\mathrm{ang}(\Gamma)=\frac{R_\mathrm{ph}(\Gamma)}{\Gamma^2 c}\sim \left(\frac{E_\star \kappa}{4\pi c^4}\right)^{1/2} \Gamma^{-\alpha/2 - 2}\beta^{-\alpha/2}
 \label{eq:t_ang_velprofile}
.\end{equation}
Assuming $\beta \sim 1$, the requirement that at $t_\mathrm{ang}(\Gamma)=(t_\gamma - t_\mathrm{GW})$ the transparent part of the ejecta have $E(>\Gamma)> E_\mathrm{iso}/\eta$ is satisfied as long as
\begin{equation}
\begin{array}{l}
 E_\star > \left[4\pi c^4(t_\gamma-t_\mathrm{GW})\right]^{-\alpha/5}\left(E_\mathrm{iso}/\eta\right)^{\alpha/5+1}=\\
 = 1.6\times 10^{53}\,\eta_{-1}^{-11/5}\,\mathrm{erg,}
\end{array}
\end{equation}
where the numerical values are given hereafter for $\alpha=6$. Since we are assuming no upper Lorentz factor cut in the energy profile, we should also invert Eq.~\ref{eq:t_ang_velprofile} to check what the Lorentz factor is at the photosphere at $t_\mathrm{ang}=1.74\,\mathrm{s}$. We obtain 
\begin{equation}
\Gamma_\mathrm{ph}(t_\mathrm{ang}) = \left(\frac{E_\star \kappa}{4\pi c^4 t_\mathrm{ang}^2}\right)^{1/(\alpha+4)}\approx 7.6\,E_{\star,53}^{1/10}
,\end{equation}
which shows that the ejecta must possess a relatively high-speed tail.

A consistency check is that the total energy in the ejecta must be a fraction of the magnetic energy developed during the merger amplification, which is $\sim 10^{51}\,\mathrm{erg}$ \citep{Giacomazzo2015}. In order to check that, we introduce a minimum velocity $\beta_\mathrm{min}$ in the velocity profile (which corresponds to the inner rarefaction wave), below which no significant energy is present, just as in \citet{DAvanzo2018}: the total ejecta energy is therefore $E_\mathrm{tot}\sim E_\star (\Gamma_\mathrm{min}\beta_\mathrm{min})^{-\alpha}$. Such minimum velocity can be related to the time of the afterglow peak, which is approximately $t_\mathrm{peak}\sim 160\,\mathrm{d}$ \citep{Alexander2018}, as follows. We assume the afterglow to be synchrotron emission from the forward shock that forms as the ejecta decelerate in the ISM. The radius of the shock is found \citep{Hotokezaka2016} by solving the energy balance equation (assuming a uniform ISM density),
\begin{equation}
\frac{4}{3}\pi R^3 n\,m_\mathrm{p} (c\beta\Gamma)^2 = E(> \Gamma)
,\end{equation}
where $n$ is the ISM number density and $m_\mathrm{p}$ is the proton mass. The afterglow luminosity keeps rising as long as slower ejecta inject their energy into the shocked region, and it therefore peaks when the ejecta with minimum velocity $\beta_\mathrm{min}$ reach the shock. Taking the observer time to be $t_\mathrm{obs}\sim R/\beta \Gamma^2 c$, we obtain
\begin{equation}
 t_\mathrm{peak}\sim \left(\frac{3 E_\star}{4\pi n m_\mathrm{p} c^5}\right)^{1/3} \beta_\mathrm{min} (\Gamma_\mathrm{min}\beta_\mathrm{min})^{-\alpha/3 - 8/3}
.\end{equation}
The total energy in the ejecta is therefore roughly
\begin{equation}
\begin{array}{l}
  E_\mathrm{tot}\sim \left[\frac{4\pi}{3} n m_\mathrm{p} c^5 t_\mathrm{peak}^{3} E_\star^{8/\alpha}\right]^{\alpha/(8+\alpha)}\approx\\
 \approx 2\times 10^{50}\,n_{-4}^{3/7}E_{\star,53}^{4/7}\,\mathrm{erg.}\\
 \end{array} 
\end{equation}

The actual energy requirement is further reduced if the outflow is not totally isotropic, but is instead ejected along the polar direction within an effective half-opening angle $\theta_0$. Since the region close to the equatorial plane of the binary is likely filled with tidal ejecta, this is actually the most likely case. If $\theta_0\sim 45^\circ$, the afterglow would be essentially unaffected, while the total energy would be lowered by a factor $(1-\cos\theta_0)^{-1}\sim 3$.

We can therefore conclude that, in the presence of a mechanism capable of extracting $\sim $10\%\ of the energy contained in the amplified magnetic field during the initial part of the merger (e.g.~reconnection close to the light cylinder, or impulsive magnetic acceleration in the low-density polar regions), and in the presence of a form of energy dissipation within the outflow (e.g.~internal shocks, or further magnetic reconnection within the outflow) able to dissipate $\sim $10\%\ of such energy, we can still interpret GRB170817A and its afterglow within the giant-flare-like scenario proposed in Paper I.

\section{Discussion}
The fact that GRB170817A has been associated to the first GW from a NS-NS merger is exciting and somewhat surprising given expectations based on the rate and geometry of SGRBs \citep[e.g.][]{Ghirlanda2016,Wanderman2014a}. Indeed, the vast majority of the GRB community seems to agree that some new ingredient must be added to the usual recipe (\ie the traditional narrow, uniformly ultrarelativistic jet with sharp borders) in order to explain this burst. In Sect. \ref{sec:time_delay} we outlined an argument against de-beamed emission from an ultrarelativistic jet seen off-axis, showing that the short time delay between GW170817 and GRB170817A would require fine-tuning of the viewing angle. This makes such a scenario unlikely, and points instead to a picture where the gamma-ray emission comes from material moving towards the observer with a relatively small Lorentz factor $\Gamma\lesssim 10$. This is well accommodated in the jet cocoon emission scenario proposed by \citet{Lazzati2016,Lazzati2017} and supported by, for example,\LEt{please replace the following semicolons with commas and "and"; i.e. normal punctuation.}~ \citet{Kasliwal2017}, \citet{Bromberg2017}, \citet{Gottlieb2017a}, \citet{Piro2017}, \citet{Margutti2018}, \citet{Lyman2018}, \citet{Troja2018} and \citet{DAvanzo2018}, but these scenarios require the presence of a jet, which may not always be produced in NS-NS mergers. In this work we showed that the alternative explanation of the gamma-ray burst and of its afterglow light curve as a sort of giant flare powered by the magnetic field amplified during the NS-NS merger, as described in Paper I, is still possible, provided that the outflow is endowed with a velocity profile. 

During the preparation of this work, a preprint was posted on arXiv by \citet{Mooley2018}, which argues that VLBI radio observations of the source show superluminal motion indicating that the external shock still moves with $\Gamma\sim 4$ around $200$ days after GW170817, thereby supporting the presence of a successful jet. If their conclusions are correct, then our proposed mechanism is likely unimportant for this source.
Nevertheless, given that the conditions to produce our giant flare could be common to all NS-NS mergers, future observations of these events will provide new testing grounds for this scenario.

\begin{acknowledgements}
We thank the anonymous referee for helping us to significantly improve the original manuscript. We also thank Albino Perego for insightful discussions and useful suggestions.
\end{acknowledgements}

\footnotesize{
\bibliographystyle{aa}
\bibliography{/home/omsharan/Dropbox/Bibliography/library}
}

\end{document}